\documentclass[11pt]{article}
\usepackage{graphicx,epsfig}
\usepackage[tbtags]{amsmath}

\usepackage{graphics,amsfonts,amssymb, euscript, cite, array}
\newcommand{\p}{\dagger}
\newcommand{\tr}[1]{\langle #1 \rangle}
\newcommand{\qq}[1]{\textquotedblleft#1\/\textquotedblright}
\newcommand{\qs}[1]{\textquoteleft#1\/\textquoteright}
\newcommand{\sm}[1]{\centering$\scriptstyle #1$}

\newcommand{\pd}[2]{\frac{\delta #1}{\delta #2}}
\newcommand{\x}[1]{{\scriptscriptstyle#1}}

\begin{document}


\title{%
Effective vertex for
$\pi^0\gamma\gamma$\setcounter{footnote}{3}\thanks{Presented by K.
K. at 5th Int. Conf. Renormalization Group 2002, Tatransk\'a
\v{S}trba, Slovakia, March 10-16, 2002 } }
\author{
K.~Kampf\setcounter{footnote}{1}\footnote{Karol.Kampf@mff.cuni.cz}, J.~Novotn\'y\footnote{Jiri.Novotny@mff.cuni.cz}\\
\small Institute of Particle and Nuclear Physics,
Charles University,\\[-0.1cm] \small V Holesovickach 2, 180 00 Prague 8
Czech Republic
}
\date{}
\maketitle


\abstract{%
The $\pi^0\gamma\gamma$ vertex is used as an explicit example of the subtleties
connected with the application of {\it equation of motion\/} within Chiral Perturbation Theory
at the order $\mathcal{O}(p^6)$.}



\section{Introduction}
\setcounter{section}{1}\setcounter{equation}{0}
We would like to study the well-known process $\pi^0 \rightarrow \gamma \gamma$
in the domain of Chiral Perturbation theory in the simple case when these two
photons are on-shell.
This calculation can be performed in accord with Bijnens's article \cite{bijnens93},
i.e. by a direct use of Feynman diagrams or by Gasser and Leutwyler functional approach \cite{gasser}.
In the first method we have to introduce the wave function renormalization
(which does not appear in the second one). We will briefly summarized this procedure
for the Chiral symmetry group $SU(3)_R \times SU(3)_L$ which is spontaneously broken to
the vector subgroup $SU(3)_V$.
\section{Three-flavour case}
Lagrangian is in the standard chiral power counting given by
\begin{equation}
\mathcal{L} = \mathcal{L}^{(2)} + \mathcal{L}^{(4)} +\mathcal{L}^{\text{WZ}}
+\mathcal{L}^{(6)} + \ldots,
\label{L}
\end{equation}
where the chiral invariant terms relevant for our process are given by
\begin{align}
\mathcal{L}^{(2)} &= \frac{F_0^2}{4}\bigl( \langle D_\mu U^\p D^\mu U \rangle
+ \langle \chi^\p U + \chi U^\p \rangle
\bigr)\\
\mathcal{L}^{(4)} &=
L_4 \langle D^\mu U^\p D_\mu U\rangle \langle \chi^\p U + \chi U^\p \rangle
+L_5 \langle D^\mu U^\p D_\mu U (\chi^\p U + U^\p \chi)\rangle + \ldots\\
\intertext{and the Wess-Zumino term which contains the anomaly}
{\cal L}^{\text{WZ}} &= -\frac{N_C e}{48\pi^2}A_\mu J^\mu + i\frac{N_C e^2}{24\pi^2}
\varepsilon^{\mu\nu\alpha\beta}\partial_\mu A_\nu A_\alpha T_\beta,
\end{align}
where we have used
\begin{align*}
U &= {\rm e}^{i \frac{\phi}{F_0}}, \qquad
\phi = \begin{pmatrix}\pi^0 + \eta/\sqrt{3} &\sqrt{2}\pi^+ & \sqrt{2}K^+\\
\sqrt{2}\pi^- & -\pi^0 + \eta/\sqrt{3} & \sqrt{2}K^0\\
\sqrt{2}K^- &\sqrt{2}\bar{K}^0 & -2/\sqrt{3} \eta
\end{pmatrix}\\
J^\mu &= \varepsilon^{\mu\nu\alpha\beta} {\rm Tr}(Q L_\nu L_\alpha L_\beta
  +Q R_\nu R_\alpha R_\beta)\\
T_\beta &= {\rm Tr}(Q^2 L_\beta +Q^2 R_\beta +\frac{1}{2}U^\p Q U Q L_\beta +\frac{1}{2} U Q U^\p
Q R_\beta)
\end{align*}
with $L_\mu=U^\p \partial_\mu U, \; R_\mu = (\partial_\mu U)U^\p$.
We will not need the explicit form of $\mathcal{O}(p^6)$ terms.

The wave function renormalization factor $\mathcal{Z}$ is the residue of the complete propagator
of the pion field at the physical mass, as a result we get
\begin{equation}
\frac{1}{\mathcal{Z_\pi}} = 1 + \frac{1}{F_0^2}
\Bigr[ -\frac{2}{3}A(m_\pi^2)-\frac{1}{3}A(m_K^2)+8 L_4 (2 m_K^2 + m_\pi^2) + 8 L_5 m_\pi^2\Bigl],
\label{Z}
\end{equation}
where
\begin{equation}
A(m^2) \equiv
\int \frac{d^d l}{(2\pi)^d}\frac{i}{l^2 - m^2} \biggr|_{d\rightarrow 4}
= 2 m^2 \lambda +\frac{m^2}{16\pi^2}\log{\frac{m^2}{\mu^2}}.
\end{equation}
The physical decay constant $F_\pi$ is defined by means of axial current through
\begin{equation}
\langle \Omega | A^a_\mu(x) |\pi^b(p)\rangle = i \delta^{ab}
F_{\pi^a} p_\mu {\rm e}^{-ip.x}
\end{equation}
The output of calculation is
\begin{equation}
F_\pi = F_0 \Bigl(1+\frac{1}{F_0^2}\bigl[-A(m_\pi^2) -\frac{1}{2}A(m_K^2) + 4L_4(2m_K^2 + m_\pi^2)
+4L_5 m_\pi^2 \bigr] \Bigr).
\label{F}
\end{equation}
$\mathcal{Z_\pi}$ and $F_\pi$ ((\ref{Z}) and (\ref{F})) are the basic ingredients needed for calculating
any amplitude involving pions which are governed by Lagrangian (\ref{L}).

The amplitude for process $\pi \rightarrow \gamma \gamma$
up to next-to-leading order without the $\mathcal{O}(p^6)$ terms is simply given by \cite{bijnens93}
\begin{equation}
A(\pi^0 \rightarrow \gamma(k) \gamma(l)) = -\frac{N_C}{3}\frac{i \alpha}{\pi F_\pi}
\varepsilon^{\mu\nu\alpha\beta} \epsilon^*_\mu (k) \epsilon^*_\nu (l) k_\alpha l_\beta,
\label{A}
\end{equation}
which differs from the lowest order only by changing $F_0$ to $F_\pi$.
We can also see that the possible $\mathcal{O}(p^6)$ contributions have to be finite.
This confirms explicitly the result which can be obtain directly using the methods
of the heat kernel expansion and dimensional regularization,
particulary the divergent part of the one-loop generating functional
relevant for this process is given by
\begin{multline}
Z_{\text{1-loop}}^{\pi\gamma\gamma} = -\frac{i}{32 \pi^2 (d-4)} \frac{N_c N_f}{72 \pi^2 F_0^2}
\varepsilon^{\mu\nu\alpha\beta} \partial^\gamma F_{\gamma\nu} F_{\alpha\beta}\\
\times \bigl(\langle QUQ\partial_\mu U^\p - QU^\p Q \partial_\mu U \rangle
 - \langle Q^2 (R_\mu + L_\mu) \rangle \bigr).
\end{multline}
\section{Two-flavour case}
As we have stated this was a standard calculation in $SU(3)$ case.
We would like to turn our attention to the chiral symmetry
$SU(2)_R \times SU(2)_L \rightarrow SU(2)_V$.
For this purpose we can use already given Lagrangian but now we take $U$ to be a $2\times 2$ matrix
\begin{equation}
U = {\rm e}^{i \frac{\phi}{F_0}}, \qquad
\phi = \begin{pmatrix}\pi^0 &\sqrt{2}\pi^+ \\
\sqrt{2}\pi^- & -\pi^0
\end{pmatrix}
\end{equation}
Using simple identities for $2\times 2$ matrix one gets for (\ref{L})
\begin{equation}
\mathcal{L}' = \mathcal{L}'^{(2)} + \mathcal{L}'^{(4)} +\mathcal{L}'^{\text{WZ}}
+\mathcal{L}'^{(6)} + \ldots
\end{equation}
where the forms of $\mathcal{L}'^{(2)}$ and $\mathcal{L}'^{\text(WZ)}$ stay unchanged
\footnote{This is true for vectorial sources or photons as can be directly check using Kaiser's $WZ$ Lagrangian \cite{kaiser01}},
whereas chiral invariant $O(p^4)$ terms are reduced to
\begin{equation}
\mathcal{L}'^{(4)} =
(2L'_4+L'_5)\tr{(d_\mu U^\p d^\mu U)(\chi^\p U + U^\p \chi)} + \ldots
\end{equation}
The number of $\mathcal{O}(p^4)$ low energy couplings (LEC) decreases from 10 to 7.

With this input we would obtain (instead of (\ref{Z}),(\ref{F}) and (\ref{A}))
\begin{equation}
\frac{1}{\mathcal{Z}'_\pi} = 1 + \frac{1}{F_0^2}
\Bigr[ -\frac{2}{3}A(m_\pi^2) + 8 m_\pi^2 (2 L'_4 + L'_5) \Bigl],
\label{Z'}
\end{equation}
\begin{equation}
F'_\pi = F'_0 \Bigl(1+\frac{1}{F_0^2}\bigl[-A(m_\pi^2) + 4 m_\pi^2 (2L'_4 + L'_5)  \bigr] \Bigr)
\label{F'}
\end{equation}
\begin{equation}
A'(\pi^0 \rightarrow \gamma(k) \gamma(l)) = -\frac{N_C}{3}\frac{i \alpha}{\pi F'_\pi}
\varepsilon^{\mu\nu\alpha\beta} \epsilon^*_\mu (k) \epsilon^*_\nu (l) k_\alpha l_\beta.
\label{A'}
\end{equation}
We have changed the notation of LECs to stress that now we are in different theory,
where of course $L_i \neq L'_i$ (renormalized).
However, the physical quantities have to stay same and so $F'_\pi = F_\pi$ and
$A' = A$. This can be used for finding the relations between $L_i$ and $L'_i$.
We have also introduced $F'_0$ which is the pion decay constant in $SU(2)$ chiral limit, i.e.
for $m_u=m_d=0$, $m_s\neq 0$.
One can calculate it from (\ref{F}) to find
\begin{equation}
F'_0 = F_0 \Bigl( 1 + \frac{1}{F_0^2}\bigl[ -\frac{1}{2} A(m_s B)
+ 8 L_4 m_s B \bigr] \Bigr)
\end{equation}
From $F'_\pi = F_\pi$ we obtain for finite part
\begin{equation}
4 ( 2 L'^r_4 + L'^r_5) = 4 (2 L^r_4 + L^r_5) - \frac{1}{2} \frac{1}{32 \pi^2} \log{\frac{m_K^2}{\mu^2}} - \frac{1}{64\pi^2}
\label{fin}
\end{equation}
and with usual definition $L_i = L^r_i + \Gamma_i \lambda$ for infinite part
\begin{align}
4 ( 2 \Gamma'_4 + \Gamma'_5) &= -\frac{1}{2} + 4 (2\Gamma_4 + \Gamma_5)
\intertext{For $\Gamma_4 = \frac{1}{8}$, $\Gamma_5 = \frac{3}{8}$ we get}
4 ( 2 \Gamma'_4 + \Gamma'_5) &= 2.
\label{infi}
\end{align}
In order to establish the canonical form of Gasser and Leutwyler $SU(2)$ Lagrangian at $\mathcal{O}(p^4)$ order \cite{gasser},
we use the equation of motion, derived from $\mathcal{O}(p^2)$ Lagrangian
\begin{equation}
\EuScript{O}_{\text{EOM}}^{(2)} \equiv d^2 U U^\p - U d^2 U^\p - \chi U^\p + U \chi^\p
+\frac{1}{2} \langle \chi U^\p - U \chi^\p \rangle
\end{equation}
and using the integrations by parts and some simple algebra,
one can rewrite $\mathcal{L}'^{(4)}$ to the equivalent form
\begin{equation}
\begin{split}
\mathcal{L}'^{(4)} = \phantom{+\ }&(2L'_4+L'_5)\tr{d_\mu U d^\mu \chi^\p + d_\mu U^\p d^\mu \chi}
-\tfrac{1}{4}(2L'_4+L'_5) \tr{\chi^\p U + U^\p \chi}^2\\
-\ & \frac{1}{2} (2 L'_4 + L'_5) \tr{\EuScript{O}^{(2)}_{\text{EOM}} (\chi U^\p - U \chi^\p)} +\ldots
\end{split}
\end{equation}
which, of course, reproduce the same result as (\ref{Z'}--\ref{A'}).
Now we can introduce a new notation
\begin{equation}
l_4 \equiv 4 (2 L'_4 + L'_5)
\label{lL}
\end{equation}
so we get
\begin{equation}
\mathcal{L}'^{(4)} = \mathcal{L}^{(4)}_{\text{can}} - \frac{1}{2} \frac{l_4}{4}
\tr{\EuScript{O}^{(2)}_{\text{EOM}} (\chi U^\p - U \chi^\p)},
\label{stX}
\end{equation}
where $\mathcal{L}^{(4)}_{\text{can}}$ is the desired canonical form.
For further purposes we will set $l_4 \rightarrow X$ in the last term of \eqref{stX}.
The canonical form, up to order $\mathcal{O}(p^4)$ is simply obtained by setting $X=0$.
A review of our calculation, up to now, is summarized in Table \ref{recap}.
\begin{table}[ht]
\setlength{\tabcolsep}{0pt}
\begin{tabular}{|m{2cm}|m{3cm}|m{3cm}|m{3cm}|m{3cm}|}
\cline{2-5}
\multicolumn{1}{l|}{} &\centering $SU(3)$ & \multicolumn{3}{c|}{$SU(2)$} \\
\cline{3-5}
\multicolumn{1}{l|}{} &&\centering $SU(3)$-like & \centering standard + $X$ & \begin{center} standard \end{center} \\
\hline%
\epsfig{figure=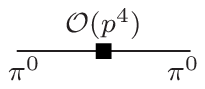} &\sm{\frac{1}{F^2}[8L_4(2m_K^2+m_\pi^2)} \sm{\phantom{a}\hspace{1cm}+8 L_5 m_\pi^2]}&
\sm{\frac{1}{F^2}8m_\pi^2(2L'_4+L'_5)}&\sm{\frac{1}{F^2}m_\pi^2 2X}& \begin{center} \sm{0} \end{center} \\ \hline
\epsfig{figure=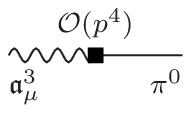} &\sm{\frac{1}{F}[8L_4(2m_K^2+m_\pi^2)} \sm{\phantom{a}\hspace{1cm}+8 L_5 m_\pi^2]}
&\sm{\frac{1}{F}8m_\pi^2(2L'_4+L'_5)}
&\sm{\frac{1}{F}m_\pi^2(l_4+X)} & \begin{center}\sm{\frac{1}{F}m_\pi^2 l_4}\end{center} \\ \hline
\sm{F_\pi}
&\sm{F_0(1+\frac{1}{F^2}[-A_\pi-\frac{A_K}{2}} \sm{+4L_4(2m_K^2+m_\pi^2)} \sm{\phantom{aaaaaaaa}+4L_5m_\pi^2])}
&\sm{F'_0(1+\frac{1}{F^2}[-A_\pi} \sm{+ 4m_\pi^2(2L'_4+L'_5)])}
&\sm{F'_0(1+\frac{1}{F^2}[\text{-}A_\pi+m_\pi^2l_4])}
&$\scriptstyle F'_0(1+\frac{1}{F^2}[\text{-}A_\pi+m_\pi^2l_4])$
\\ \hline
\end{tabular}
\caption{Recapitulation of calculation of $F_\pi$}
\label{recap}
\end{table}

By means of relation \eqref{lL} we can verify the relation between the standard $SU(3)$ and $SU(2)$ chiral expansion
(cf.(\ref{fin}-\ref{infi}))
\begin{equation}
l^r_4 = 4(2 L^r_4 + L^r_5) - \frac{1}{64\pi^2} \log{\frac{m_\pi^2}{\mu^2}} - \frac{1}{64\pi^2}
\end{equation}
and
\begin{equation}
\gamma_4 = 2
\end{equation}
with the definition $l_i = l^r_i + \gamma_i \lambda$.

However, it rests to show, what has allowed us to set $X=0$
and so pass over to canonical form of Lagrangian in \eqref{stX}.
Before doing that, let us stress that the physically relevant quantities,
calculated up to order $p^4$, should not depend on the choice of the Lagrangian
$\mathcal{L}^{(4)}$ (i.e. on the fact whether we set $X=0$ or not)
and this is, indeed, the case of $F_\pi$ (see Table.\ref{recap}),
but for the amplitude $A$ (which is of order $\mathcal{O}(p^6)$)
we have in \qs{standard + $X$} case:
\begin{equation}
A^X(\pi^0 \rightarrow \gamma(k) \gamma(l)) = -\frac{N_C}{3}\frac{i \alpha}{\pi F_\pi}
\varepsilon^{\mu\nu\alpha\beta} \epsilon^*_\mu (k) \epsilon^*_\nu (l) k_\alpha l_\beta
\Bigl(1 + \frac{1}{F} m_\pi^2 (l_4 -X)\Bigr)
\end{equation}
which is $UV$-safe, at the level of $\mathcal{O}(p^4)$ Lagrangians, only if
$X^{div}=l_4^{div}$, e.g. in the case when we take {\it full\/} form \eqref{stX}.
In the standard Leutwyler-Gasser case of $SU(2)$ Lagrangian it follows that
this necessitates the additional (not only finite) term of the $\mathcal{O}(p^6)$ order.
We can naively provide it using the equation of motion.
\section{General case}
Formula (\ref{stX}) is a motivation for studying the general case of the supplemental non-canonical term
$\tr{\EuScript{O}^{(2)}_\text{EOM}\mathcal{F}}$ in the action
\begin{equation}
S = S^{(2)} + S^{(4)}_\text{LG} + \int\tr{\EuScript{O}^{(2)}_\text{EOM}\mathcal{F}} + S^{(6)} + \ldots,
\end{equation}
where $S^{(4)}_\text{LG}$ is the Leutwyler and Gasser canonical form.\\
Full equation of motions could be obtained from
\begin{equation}
\EuScript{O}_\text{EOM} = \frac{4}{F^2} \Bigl( U \pd{S}{U} - \pd{S}{U^\p} U^\p
-\frac{1}{2}\tr{U\pd{S}{U} - \pd{S}{U^\p} U^\p} \Bigr)
\end{equation}
which can be decomposed
\begin{equation}
\EuScript{O}_\text{EOM} =\EuScript{O}^{(2)}_\text{EOM} + \delta\EuScript{O}^{(4)}_\text{EOM} + \ldots
\end{equation}
Naively we set $\EuScript{O}_\text{EOM} = 0$ and so find
\begin{equation}
S = S^{(2)} + S^{(4)}_\text{LG} + S^{(6)} - \int\tr{\delta \EuScript{O}^{(4)}\mathcal{F}} + \ldots
\label{Snaiv}
\end{equation}
However, this application of the equation of motions is not correct and may be dangerous at the order
$\mathcal{O}(p^6)$ and the correct procedure is the following change of variable
\begin{equation}
U \rightarrow {\rm e}^{-\mathcal{F}} U = U + \sum_{n=1}^\infty \delta^{(n)} U
\qquad \text{with}\quad
\delta^{(n)}U \equiv \frac{(-1)^n}{n!} \mathcal{F}^n U,
\end{equation}
as was pointed out in \cite{scherer}.\\
We assume $\mathcal{F}=\mathcal{O}(p^2)$ and we get finally
\begin{equation}
\begin{split}
S' = &S^{(2)} + S^{(4)}_\text{LG} + S^{(6)} - \int\tr{\delta \EuScript{O}^{(4)}_\text{EOM} \mathcal{F}}\\
&+\frac{1}{2}\int\tr{(U\pd{S^{(2)}}{U} + \pd{S^{(2)}}{U^\p}U^\p)\mathcal{F}^2}
+\frac{F_0^2}{4}\int\tr{d_\mu(U^\p \mathcal{F})d^\mu(\mathcal{F}U)} \\
&-\iint\Bigl(\pd{S^{(2)}}{U_\x{kl}}\delta U_\x{ij}\pd{\delta^{(1)}U_\x{kl}}{U_\x{ij}} +
\pd{S^{(2)}}{U_\x{kl}^\p}\delta U_\x{ij}\pd{\delta^{(1)}U^\p_\x{kl}}{U_\x{ij}} \\
&\phantom{-\iint\Bigl(}+\pd{S^{(2)}}{U_\x{kl}}\delta U_\x{ij}^\p\pd{\delta^{(1)}U_\x{kl}}{U_\x{ij}^\p}
+\pd{S^{(2)}}{U_\x{kl}^\p}\delta U_\x{ij}^\p\pd{\delta^{(1)}U_\x{kl}^\p}{U_\x{ij}^\p}
\Bigr) + \mathcal{O}(p^8)
\end{split}
\label{Scor}
\end{equation}
We can see that this differ from the naive calculation \eqref{Snaiv} by supplementary terms.
However, this terms (in the second, third and fourth line of \eqref{Scor})
only depend on $S^{(2)}$ and $\mathcal{F}$, so they are of even intrinsic parity and therefore they do not contribute
to the process $\pi^0 \rightarrow \gamma\gamma$.

We confirm a naive prediction of the infinite counterterm of $\mathcal{O}(p^6)$ order which is
\begin{equation}
- \int\tr{\delta \EuScript{O}^{(4)}_\text{EOM} \mathcal{F}} \longrightarrow
\frac{N_c\alpha}{6\pi}\frac{l_4}{F_0^3}m_\pi^2 \varepsilon^{\mu\nu\rho\sigma}
\partial_\mu A_\nu \partial_\rho A_\sigma \pi^0
\end{equation}
and thus we get finally
\begin{equation}
A^{SU(2)}(\pi^0 \rightarrow \gamma(k) \gamma(l)) = -\frac{N_C}{3}\frac{i \alpha}{\pi F_\pi}
\varepsilon^{\mu\nu\alpha\beta} \epsilon^*_\mu (k) \epsilon^*_\nu (l) k_\alpha l_\beta
\end{equation}
(up to finite $\mathcal{O}(p^6)$ counterterms).

\section{Conclusion}
We find there are two types of $\mathcal{O}(p^6)$ counterterm:
first type, which is independent of the form of $\mathcal{O}(p^4)$ Lagrangian
(see eg.\cite{bijnens01}) and the second type, which is dependent, as we have explicitly showed
in the case of $\pi^0\gamma\gamma$ amplitude.
Of course, the physical quantity up to order $\mathcal{O}(p^4)$ (e.g. $F_\pi$ in our text) stay unchanged.
\\[0.5cm]
{\bf Note added.\ }{\it After this work had been finished and
presented, paper \cite{Ananthanarayan02} appeared in which similar
problems were discussed.}
\\[0.5cm]
{\bf Acknowledgement.\ } This work was supported by the program
\qq{Research Centres} project number LN00A006 of the Ministry of
Education, Youth and Sport of the Czech republic.

\end{document}